\documentclass[conference]{IEEEtran}
\IEEEoverridecommandlockouts
\usepackage{cite}
\usepackage{amsmath,amssymb,amsfonts}
\usepackage{algorithmic}
\usepackage{textcomp}
\usepackage[table]{xcolor}
\usepackage{graphicx}
\usepackage{multirow}
\usepackage[hyphens]{url}
\usepackage{hyperref}
\usepackage{tikz}
\usepackage{subfig,xspace}
\newcommand*\circled[1]{\tikz[baseline=(char.base)]{\node[shape=circle,fill,inner sep=0pt,minimum size=1pt] (char) {\textcolor{white}{#1}};}}
\usepackage[normalem]{ulem}
\def\BibTeX{{\rm B\kern-.05em{\sc i\kern-.025em b}\kern-.08em
    T\kern-.1667em\lower.7ex\hbox{E}\kern-.125emX}}

\newcommand{\exwarp}{{\em ExWarp}\xspace}

\begin{document}

\title{\textit{ExWarp}: Extrapolation and Warping-based Temporal Supersampling for High-frequency Displays}
\author{\IEEEauthorblockN{ Akanksha Dixit} 
    \IEEEauthorblockA{\textit{Electrical Engineering} \\
    \textit{Indian Institute of Technology}\\
    New Delhi, India \\
    Akanksha.Dixit@ee.iitd.ac.in} 
    \and
    \IEEEauthorblockN{Yashashwee Chakrabarty} 
    \IEEEauthorblockA{\textit{Computer Science and Engineering} \\
    \textit{Indian Institute of Technology}\\
    New Delhi, India \\
    mcs222057@cse.iitd.ac.in}
    \and
    \IEEEauthorblockN{Smruti R. Sarangi} 
    \IEEEauthorblockA{\textit{Electrical Engineering} \\
    \textit{Indian Institute of Technology}\\
    New Delhi, India \\
    srsarangi@cse.iitd.ac.in}
   }
   
\maketitle

\begin{abstract}
High-frequency displays are gaining immense popularity because of their increasing use in video games and virtual
reality applications.  However, the issue is that the underlying GPUs cannot continuously generate frames at this high
rate -- this results in a less smooth and responsive experience. Furthermore, if the frame rate is not synchronized with the refresh
rate, the user may experience screen tearing and stuttering. Previous works propose increasing the frame rate to provide
a smooth experience on modern displays by predicting new frames based on past or future frames. Interpolation and
extrapolation are two widely used algorithms that predict new frames.  Interpolation requires waiting for the future
frame to make a prediction, which adds additional latency.  On the other hand, extrapolation provides a better quality
of experience because it relies solely on past frames --  it does not incur any additional latency. The simplest method
to extrapolate a frame is to warp the previous frame using motion vectors; however, the warped frame may contain
improperly rendered visual
artifacts due to dynamic objects -- this makes it very challenging to design such a scheme. Past work has used DNNs
to get good accuracy, however, these approaches are slow.
This paper proposes \exwarp -- an approach based on
reinforcement learning (RL) to intelligently choose between the slower DNN-based extrapolation and faster warping-based
methods to increase the frame rate
by 4$\times$ with an almost negligible reduction in the perceived image quality.
\end{abstract}


\pagestyle{plain}

\section{Introduction}
\label{sec:Introduction}

The CAGR (compound annual growth rate) for the global gaming market is projected to be 7.7\% over the next 5
years~\cite{2022Gaming}.  This will lead to a total revenue of roughly 532.97 billion USD by 2027~\cite{2021Games}. To
provide as realistic an experience as possible, displays are supporting increasingly higher refresh rates. We have moved
from 30 Hz to 120 Hz over the last few years. Such systems help players feel fully involved in the virtual world. Given
that the human vision system is exceptionally sensitive and can sometimes detect a lag as low as 2
ms~\cite{latencyforhuman}, gamers prefer ultra-high refresh rates. For newer
head-mounted displays, latencies greater than 7 ms may result in
motion sickness and dizziness~\cite{latencyinvr}. Even for non-gamers, research has shown that an inter-frame
duration of 25 ms~\cite{latencyforperf, jota2013fast} can cause issues. They will perceive a {\em high
latency} in interactive tasks.   The only solution is to render and display frames as fast as possible.

To meet these latency requirements, display vendors have launched monitors with high refresh rates such as 120 Hz, 240
Hz and 360 Hz displays~\cite{360Nvidia,360}. Recently, Dell launched the Alienware 500Hz Gaming
Monitor. Mobile companies are also incorporating such displays into their devices.~\cite{Smartphones}. Similarly, there
are GPUs such as the NVIDIA RTX series GPUs, which can render up to 360 frames per second at 1080p resolution~\cite{gpus}.
Still users are not guaranteed to have a seamless experience because the actual frame rate depends
upon various parameters such as the frame resolution, frame complexity, etc., and it varies a lot during the application
Furthermore, if the frame rate is not synchronized with the refresh rate of the
monitor, the user may experience glitches known as screen tearing and screen stuttering~\cite{tearing_stuttering}. That
is why there exist synchronization algorithms such as G-Sync and Free-Sync~\cite{gsync_freesync} in NVIDIA GPUs
that modulate the
refresh rate to synchronize it with the frame rate and prevent screen tearing and screen stuttering. However, this is
not an ideal solution. Let us assume we have a 144 Hz monitor but the GPU is only supplying frames at 45 fps, then these
synchronization techniques reduce the refresh rate to 45 Hz, which results in an ineffective use of the monitor's
capabilities. Hence, this work aims to {\em temporally} supersample the frames for enabling the use of high refresh rate
displays.

Recent works propose two ways to increase the frame rate: spatial
supersampling~\cite{herzog2010spatio,nehab2007accelerating,yang2008geometry} and temporal
supersampling~\cite{andreev2010real, bowles2012iterative, extranet,yang2011image}. These works exploit the fact that
with the increase in image resolution, there exists a similarity between neighboring pixels in spatial and temporal
domains known as spatial and temporal coherence~\cite{herzog2010spatio}. Spatial supersampling increases the frame rate
by rendering the frames at a much lower resolution and then increases the resolution of the rendered frame before
displaying them using interpolation. On the other hand, temporal supersampling generates entirely new frames on the fly
using an already rendered frame. For temporal supersampling, there are two popular methods: \textit{Interpolation} and
\textit{Extrapolation}~\cite{extranet}. Most of the existing works including NVIDIA's latest supersampling method, DLSS
3 (Deep Learning SuperSampling)~\cite{DLSS3}, use optical flow-based interpolation to generate new frames.  The problem
with the interpolation method is that one needs to wait till the next frame is rendered to start the interpolation
(figure out all frames in between). 
This introduces an unnecessary delay leading to an increased input latency (refer to
Section~\ref{subsubsec:inter_vs_extra}), which degrades the overall performance. 
The advantage of these methods is that they cover occlusion and disocclusion steps. 

ExtraNet~\cite{extranet}, an extrapolation method for temporal supersampling, proposes a
way that does not rely on optical flow and extrapolates the new frame solely based on the past few frames. To handle
occlusion and dynamic objects, ExtraNet uses a few intermediate buffers that are generated during the rendering
process. It relies on a complex neural network to do this task; hence it is slow. Due to its significant latency, it
upsamples the frame rate only by 1.5 to 2$\times$. This work proposes \exwarp --  a faster extrapolation-based method to upsample
the frame rate further for high-frequency displays: from 30 to 120 Hz.  Our primary contributions are:\\ \circled{1} We show that two
widely used methods used for predicting frames -- warping and extrapolation -- can be combined for temporal supersampling in
real-time. \\ \circled{2} We identify a few features that define the current state of the scene, i.e., the presence of
dynamic objects and camera movement.\\ \circled{3} We propose a reinforcement learning (RL) based approach that uses
these identified features to intelligently choose between extrapolation and warping based methods to increase the frame rate
by almost 4$\times$ with an almost negligible reduction in the perceived image quality.  \\ \circled{4} We are able to
supersample the frame rate by nearly 4$\times$. We record an 18.02\% increase in the PSNR and a 6.58\% increase in SSIM
as compared to the state-of-the-art baseline, ExtraNet.

The paper is organized as follows. Section \ref{sec:Background} provides the background of VR architectures and
ML-based models.  Section \ref{sec:Characterization} shows the characterization of benchmarks. The implementation details are given
in Section \ref{sec:Implementation}. Section \ref{sec:Evaluation} shows the experimental results. We discuss related
work in Section \ref{sec:RelatedWork} and finally conclude in Section \ref{sec:Conclusion}.  

\section{Background}
\label{sec:Background}

\subsection{Temporal Supersampling}
\label{subsec:temporal}
As mentioned in Section~\ref{sec:Introduction}, temporal super-sampling relies on the fact that most of the content
remains the same from frame to frame. A significant portion of a frame corresponds to at least a portion of either
the previous, future, or both the frames~\cite{TemporalSuper}. This correspondence can be find out using optical flow
vectors that describe the velocities of pixels within a frame~\cite{opticalflow}.

\subsubsection{Interpolation Vs Extrapolation}
\label{subsubsec:inter_vs_extra}

\begin{figure}[!htbp]
	\centering
	\includegraphics[width=0.99\columnwidth]{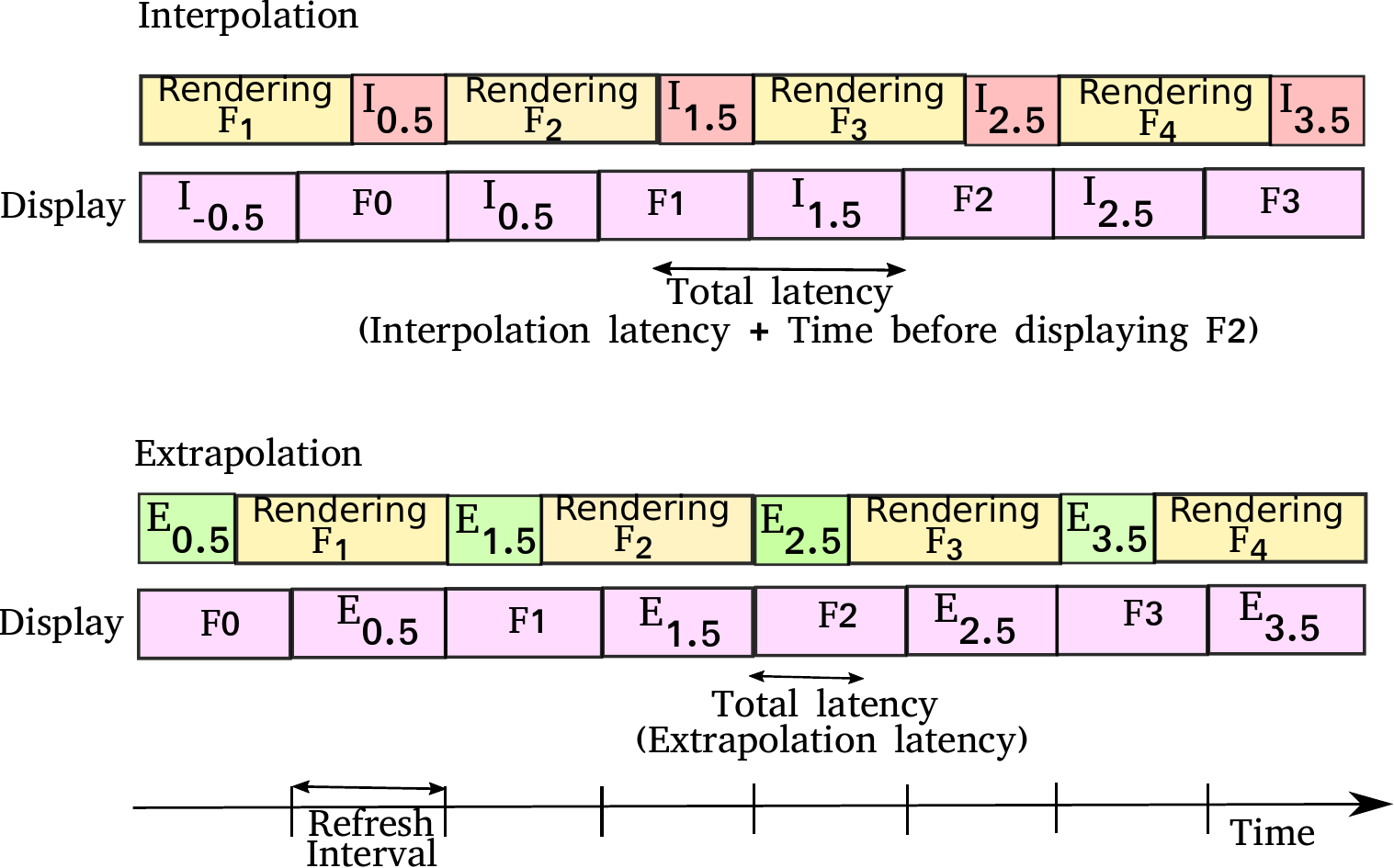}
	\caption{Interpolation and extrapolation explained. 
$F_1$, $F_2$, $F_3$ and $F_4$ are frames. $I$ and $E$ stand for interpolation and extrapolation, respectively.}
	\label{inter_vs_extra}
\vspace{-6mm}
\end{figure}

As the name suggests, {\em interpolation} predicts a frame in between two already rendered neighboring frames. Whereas,
extrapolation predicts frames based on the past frame(s) without considering future frame(s).
Figure~\ref{inter_vs_extra} explains these two algorithms in detail. In the figure, we observe that both interpolation
and extrapolation introduce some latency in the system, which is their own operational latency. Both processes double the
frame rate by generating a new frame after each rendered frame and display the frame in the following order $0, 0.5, 1,
1.5, 2, ..$. However, interpolation introduces an additional latency. Let us consider the newly generated frame with
suffix $1.5$. In interpolation, the frame $I_{1.5}$  is generated using two frames $F_1$ and $F_2$. Since frame
$I_{1.5}$ needs to be displayed before $F_2$, it waits for $F_2$, holds it, starts interpolating $I_{0.5}$, and first
displays the interpolated frame, and then $F_2$. Hence, the input latency becomes interpolation cost + the time before
displaying $F_2$. Whereas in the case of extrapolation, the new frame, $E_{1.5}$ is generated only based on the past frame
$F_1$ and all the frames are displayed at the very next refresh cycle.

\subsection{Image Warping}
\label{warping}
Image warping is a reprojection technique that maps all locations in one image to locations in a second image. It can be
used to distort the original image in a way that serves a certain purpose. It can be used to perform various tasks such
as correcting image distortion as well as for creative purposes like morphing ~\cite{beier1992feature}. One such task is
frame prediction by warping the current frame to predict a future frame ~\cite{Niklaus_2020_CVPR}. The accuracy of using
warping on frame prediction depends on how well we understand the motion between the two frames. Most modern approaches
use machine learning and Deep Neural Networks (DNNs) to estimate this motion ~\cite{jin2023enhanced} ~\cite{extranet}.

\subsection{Reinforcement Learning}
\label{RL}
Reinforcement Learning (RL) ~\cite{sutton1999reinforcement} is a machine learning-based technique that uses information about
the environment and the feedback from its actions to learn an action inside the environment. It has been derived from
reinforcement theory ~\cite{luthans1999reinforce}, which argues that human behavior is a direct result of the
consequences of one's actions. In machine learning, this technique generally doesn't require any labeled data but
requires the problem to be formulated as an actor in an environment defined by a tuple of the state space, action space, and
associated rewards. The state space defines all the legal states for the actor to be in; however, note that most
RL problems operating using incomplete information -- the state space does not capture all aspects of the environment
completely . The {\em action space} is the
collection of all the actions that the actor is allowed to take in a state. The rewards are the gains/loss associated
with each action in a particular state.

\section{Characterization}
\label{sec:Characterization}
In this section, we first show the workloads used for experiments and the platform configuration for running
experiments. As mentioned in Section~\ref{subsubsec:inter_vs_extra}, interpolation adds latency to the system in
addition to its inherent operating latency, hence  extrapolation is a preferable choice for temporal supersampling in
real-time rendering systems. However, even extrapolation has some latency. So, in this section, we show the latency of
the various steps involved in the extrapolation process used in ExtraNet. This is to find the reasons for its
unacceptably large latency and possible solutions.

\subsection{Dataset}
Similar to prior work~\cite{extranet}, we use five different applications from the Unreal Engine
marketplace~\cite{Marketplace} with different artistic backgrounds and different levels of complexities~\ref{fig:bench}.
Together they cover a range of different shading effects and transforms. Each application has scenes with 
dynamically moving objects and different inter-frame variations. The experiments are run
on an NVIDIA RTX series GPU. The detailed configuration is given in Table~\ref{tab:config}. 

\begin{table}[]
\footnotesize
\begin{center}
	
\resizebox{0.99\columnwidth}{!}{
  \begin{tabular}{|l|l|l|l|l|}
    \hline
   {\textbf{Abbr.}} &{ \textbf{Name}} &  {\textbf{Resolution}} & \textbf{API} & \textbf{Platform}    \\
    
    \hline
  \textit{LB} & Lab ~\cite{lab} & 480p & DX12 & UE   \\
    \hline

  \textit{TR} &  Tropical ~\cite{tropical} & 480p & DX12 & UE   \\
    \hline
    
  \textit{VL} &  Village ~\cite{village} & 480p & DX12 & UE   \\
    \hline
  \textit{TN} & Town ~\cite{town} & 480p & DX12 & UE  \\
   
 \textit{TN2}  &  & 720p &  &    \\
    
 \textit{TN3}  &  & 1080p &  &    \\
     \hline
  \textit{SL} &  Slum ~\cite{slum} & 480p & DX12 &UE   \\
    
\textit{SL2}  &  & 720p & DX12 &    \\
    
 \textit{SL3}  &  & 1080p & DX12 &    \\
    \hline
    \multicolumn{5}{|l|}{UE: Unreal Engine, DX: DirectX} \\
    \hline
  \end{tabular}
  }
 \end{center}
 \caption{Graphics benchmarks \label{fig:bench}}
\vspace{-3mm}
\end{table}

\begin{table}[]
\footnotesize
\begin{center}
\begin{tabular}{| l l l|} 

\hline
\rowcolor{gray}
\textbf{Parameter} &
\multicolumn{2}{l|}{\textbf{Type/Value}} \\ 
\hline\hline
CPU & \multicolumn{2}{l|}{Intel\textregistered Xeon\textregistered Gold 6226R @ 2.90GHz }\\
GPU &  \multicolumn{2}{l|}{NVIDIA RTX\texttrademark A4000 }\\ 
GPU memory &  \multicolumn{2}{l|}{16 GB}\\ 
\hline
\end{tabular}
\end{center}
\caption{Platform Configuration \label{tab:config}}
\vspace{-6mm}
\end{table}

\subsection{Extrapolation Latency}
\label{sec:extra_lat}
To the best of our knowledge, 
there is only one prominent state-of-the-art work that uses extrapolation for temporal supersampling in real-time namely 
\textit{ExtraNet}\cite{extranet}. 
We consider ExtraNet as the baseline for our work. ExtraNet is a DNN-based approach to extrapolate
frames. The authors of this work divide the extrapolation task into two stages. First, they simply warp the past frame
and then feed the warped frame to the proposed neural network to synthesize the final frame. According to them, the
warped frame created using only past frames may have some visible artifacts such as improper shadows and ghosting
effects if there are dynamic objects or there is a movement in the camera. To remove these artifacts, they first mark
invalid pixels in the warped frame and then use extrapolation to correct those pixels with the help of the neural
network. The neural network takes the last three frames into account to capture more information about the scene. 
To
mark invalid pixels as {\em holes}, they use a few intermediate buffers that are created during the rendering process, also
known as geometry buffers or G-buffers. The input to the neural network is the warped frame based on the last three rendered
frames, the corresponding images marked with holes and G-buffers. 

Hence, the steps involved in the extrapolation process are G-buffer generation, image warping, hole marking and
DNN-based inference for extrapolation. We measure the latency of each step separately for each application at different
resolutions. The results are shown in Table~\ref{tab:runtime}. These results are collected for 1000 frames per benchmark. We
make the following observations from the table:

\circled{1} The latency of all the steps except G-buffer generation is almost constant across applications for a given
resolution because it depends upon the size of the input frames. Also, the latency increases with an increase in the
resolution or image size. The latency of G-buffer generation varies across applications because it depends upon the
scene complexity.\\ 

\circled{2} For all applications, the most time-consuming step is the inference part (latency: 3.5
ms to 13.8 ms), which puts a limit on the number of frames that can be extrapolated before the actual rendered frame.
Hence, we propose to perform the inference or extrapolation only when it is necessary. We, instead, replace it with
warping,
which is faster (max latency: 4.6 ms) at the cost of accuracy.

\begin{table}[!h]
\small
\begin{center}
	
\resizebox{0.99\columnwidth}{!}{
  \begin{tabular}{|l|l|l|l|l|}
    \hline
   \multirow{2}{*}{\textbf{App.}} &  {\textbf{G-buffer}} & {\textbf{Warping}} & \textbf{Hole} & {\textbf{Network}}  \\ 
  & {\textbf{generation}} &  & \textbf{marking}& {\textbf{inference}}  \\

    \hline
    LB & 0.17 & 0.95 & 1.94 & 3.67   \\
    \hline
    TR & 0.36 &  0.89 & 1.89 & 3.78   \\
    \hline
    VL & 0.48 &  0.83 & 1.81 & 3.45   \\
    \hline
   
    TN & 0.34& 0.96 & 1.89  &  3.61  \\
    
  TN2  & 1.01 & 1.58 & 2.49 &  7.04   \\
    
  TN3  & 1.02 & 2.89 & 4.57 & 13.54   \\
    \hline
     SL & 0.24 & 0.95 & 1.93 & 3.55   \\
    
   SL2  & 1.24 & 1.67 & 2.59 &  7.09   \\
   
  SL3  & 2.1 & 2.91 & 4.63 & 13.78   \\
    \hline
  \end{tabular}
  }
 \end{center}
 \caption{Runtime (ms) breakdown of the \textit{ExtraNet} model}
\label{tab:runtime}
\vspace{-6mm}
\end{table}

\subsection{Holes in Warped Frames}
\label{holes}
\begin{figure}[!htb]
	\centering
	\includegraphics[width=0.99\columnwidth]{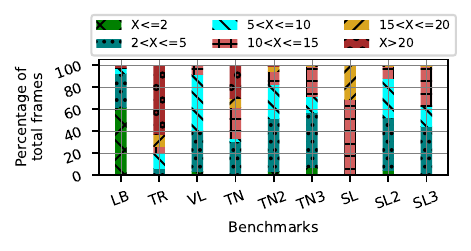}
	\caption{Percentage of holes or invalid pixels in the warped frame (denoted by $X$\%)}
	\label{hole_plot}
\vspace{-6mm}
\end{figure}

In the previous section, we discussed that ExtraNet finds invalid pixels or holes in the warped frame and then uses a
neural network to fill those holes. To see whether we can skip this hole-filling process and display the warped image
itself on the display or if this hole-filling is indeed necessary, we plot the number of holes present in warped frames
across benchmarks. We use 1000 frames for each benchmark to plot the results. The results are shown in
Figure~\ref{hole_plot}. 

We make the following observations from the figure:\\
\circled{1} The number of invalid pixels in the warped frame depends on the scenes getting rendered. There may also be
frames with no holes.\\
\circled{2} For example, in the case of LB, almost 90\% of full frames have less than 10\%
invalid pixels, whereas, for TR, more than 60\% of  total frames have more than 20\% invalid pixels.\\
\circled{3} This clearly shows that warping may provide better quality for some frames or many frames depending on the
type of application. Warping is clearly a much faster process.  
This insight motivates us to propose a method to
choose between warping and extrapolation based on the state of the current scene.

\subsection{State Representation}
\label{staterep}
As mentioned in Section~\ref{holes}, there may or may not be holes in the warped frame. If there are holes in the frame,
it means that there are dynamic objects or the camera is moving~\cite{extranet}. According to Scherzer et al.
~\cite{scherzer2011survey}, the next frame may be predicted given the previously rendered frame and motion vectors only
using warping if there are no dynamic objects in the scene. Therefore, we propose a method, a decision predictor, that
chooses between the two alternatives: warping and extrapolation. Since the performance of these two methods depends on the
scene's type or current state, we must discover a way to determine the scene's state before designing the predictor,
i.e., whether there is any movement (beyond a threshold) 
in the objects or camera leading to holes in the warped frame. Once we know the
state, we can choose between extrapolation and warping using the state information. Unlike ExtraNet, we do not require
the precise location of the holes in this particular scenario. We wish to determine whether the current state may result
in invalid pixels in the warped frame. Since we intend to produce frames for high-frequency displays, we need to find
this information quickly. Therefore, we propose a few features that capture motion information and represent the
system's current state. We use a few auxiliary buffers used in the rendering process for this purpose.

\begin{figure}[!htb]
\subfloat[Stencil]{
  \includegraphics[width=0.30\columnwidth]{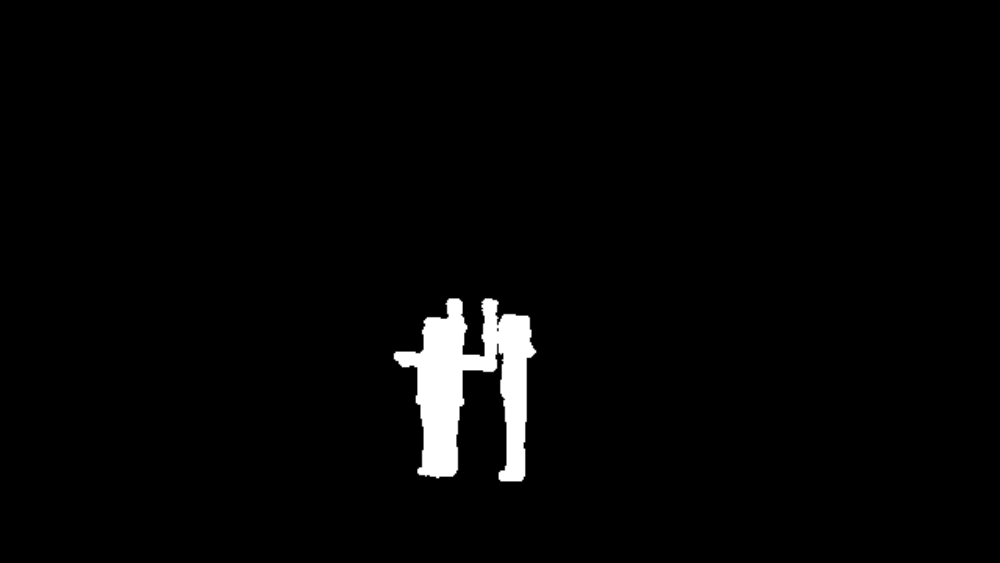}
}
\subfloat[World Normal]{
  \includegraphics[width=0.30\columnwidth]{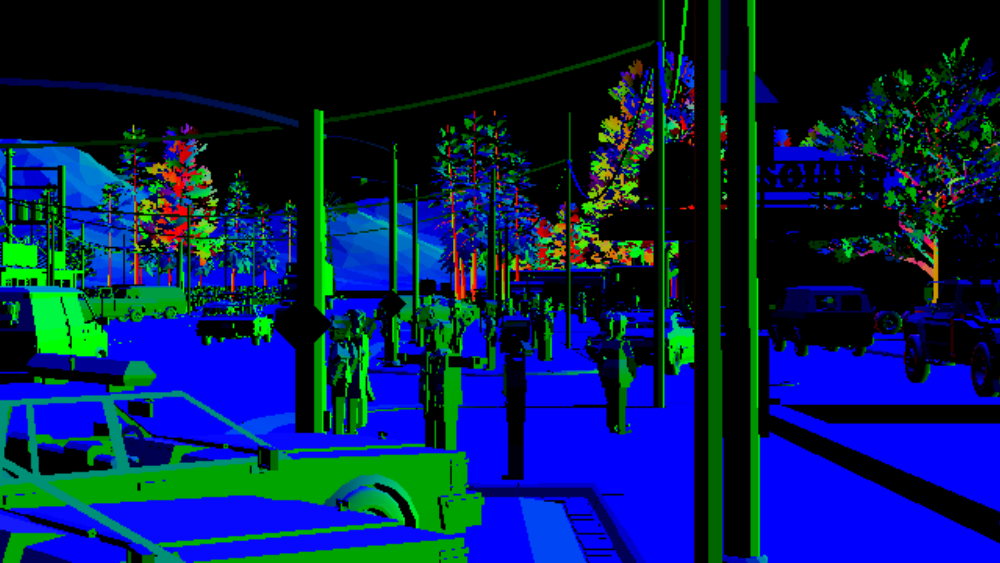}
}
\subfloat[World Position]{
  \includegraphics[width=0.30\columnwidth]{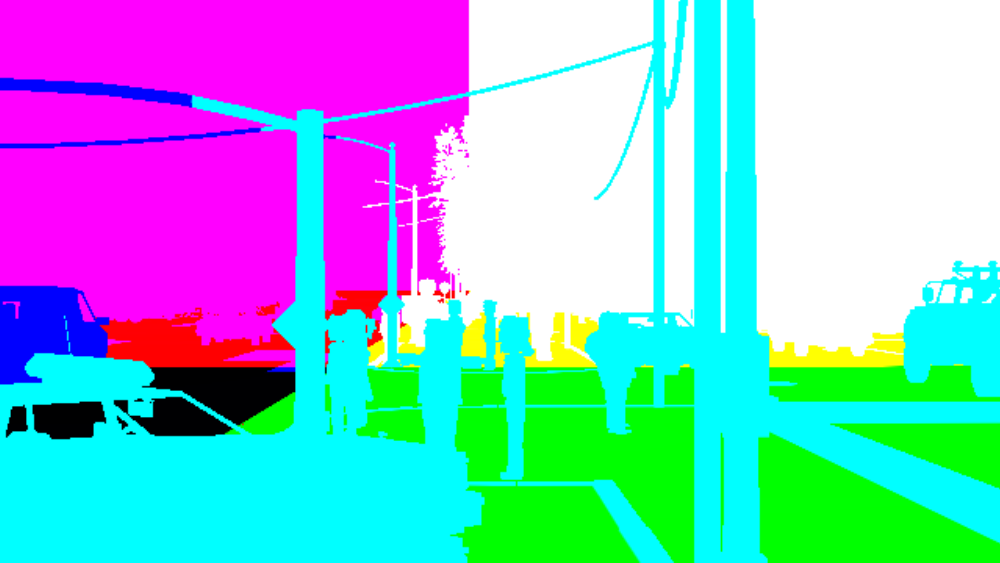}
}
\caption{Intermediate buffers used in rendering}
\label{buffers}
\vspace{-6mm}
\end{figure}

First, to capture dynamic objects, we use a motion vector buffer. A motion vector stores the motion information --
direction and magnitude -- of small blocks (areas of $16\times 16$ pixels) in the frame. Since the motion information in
the block containing the dynamic object would differ from the background or static objects, we can use this for defining
the state. We choose the variance in the motion as our feature. According to Guo et al.~\cite{extranet}, three more
buffers capture the dynamic movement information. Those are \textit{custom stencil}, \textit{world position}, and
\textit{world normal} (refer to Figure~\ref{buffers}). As clearly shown in the figure, the custom stencil buffer
directly captures dynamic objects. We use the clustering algorithm on the stencil buffer to find the number of dynamic
objects present in the frame. Next, we have the world normal and world position buffers. The values change from the last
frame to the current frame for these two buffers. We use a metric known as the Earth Movers Distance (EMD)
~\cite{rubner2000earth} to capture the change in values for these buffers. The final list of features to define the
current state is thus shown in Table~\ref{features}.

\begin{table}[!h]
\footnotesize
\begin{center}
\resizebox{0.99\columnwidth}{!}{
\begin{tabular}{|l|p{34mm}|l|} 

\hline
\textbf{Feature} &
\textbf{Description} &
\textbf{Source buffer} 
\\ 
\hline\hline

\textit{Var} & Variance in motion vector & Motion vector\\ 
\hline
\textit{$EMD_{W_N}$} & EMD between buffers corresponding to $F_t$ and $F_{t-1}$ & World normal\\
\hline
\textit{$EMD_{W_P}$} & EMD between buffers corresponding to $F_t$ and $F_{t-1}$ & World position\\
\hline
\textit{$N_D$} & Number of dynamic objects & Custom stencil\\
\hline
\end{tabular}
}
\end{center}
\caption{List of features}
\label{features}
\vspace{-6mm}
\end{table}

\subsection{Effect of the Identified Features}
\label{feature}
\begin{figure}[!htb]
	\centering
	\includegraphics[width=0.99\columnwidth]{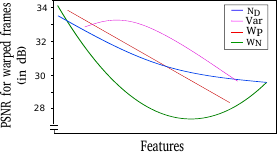}
	\caption{Effect of features on the performance of warping}
	\label{fea_var}
\vspace{-6mm}
\end{figure}

As mentioned in Section~\ref{staterep}, we use the identified features as inputs to the proposed predictor. We plot the
correlation between these variables and warping to demonstrate how they could help with the prediction. The results are
displayed in Figure~\ref{fea_var}. These results are for 1000 frames per benchmark.
Figure~\ref{fea_var} shows the variation in the quality of the warped frame. The major insights from the results are as
follows:\\ \circled{1} The pattern for all the features is the same i.e., there is a decrease in the PSNR with increase
in the
features' values.\\ \circled{2} We use this relation to design our model for the prediction.

\section{Implementation}
\label{sec:Implementation}

\subsection{Overview}
\label{sec:overview}

\begin{figure}[!htb]
	\centering
	\includegraphics[width=0.99\columnwidth]{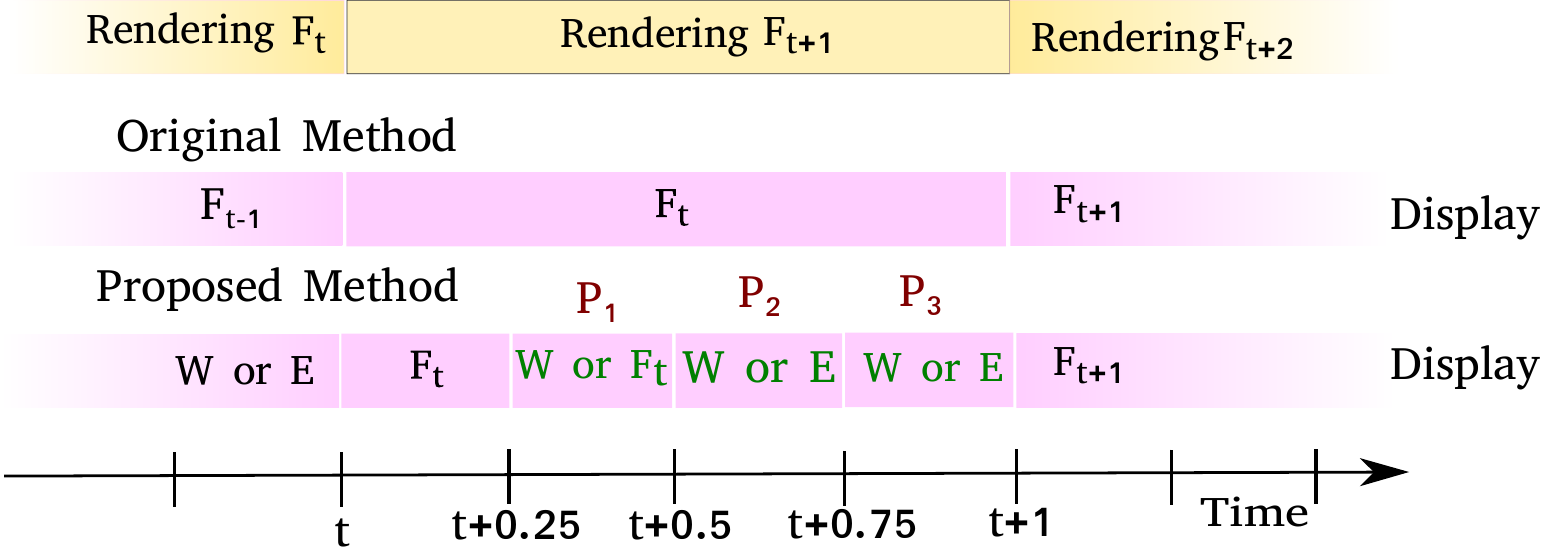}
	\caption{Overview of the proposed system. $F_t$ and $F_{t+1}$ are two rendered frames. 
$W$ and $E$ stand for warping and extrapolation, respectively. $P_1$, $P_2$, and $P_3$ are predicted frames.}
	\label{fig_overview}
\vspace{-5mm}
\end{figure}

We propose to insert three new frames at time instances $t+0.25$, $t+0.5$, $t+0.75$ between any two consecutive frames
$F_t$ and $F_{t+1}$. As mentioned in Section~\ref{sec:Introduction}, warping and extrapolation are two options for
synthesizing these new frames. Based on these two algorithms, multiple scenarios are possible (refer to
Figure~\ref{fig_overview}).  


\subsection{Problem Formulation}
\label{problem}

\begin{figure}[!htb]
	\centering
	\includegraphics[width=0.99\columnwidth]{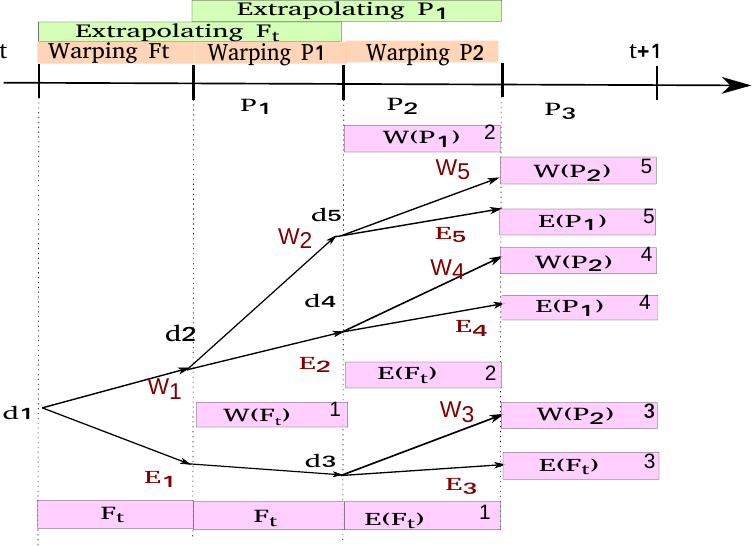}
	\caption{Flow of the proposed approach. $d1$, $d2$, $d_3$, $d_4$,and $d5$ are the 
five possible decision nodes at time $t$, $t+0.25$, and $t+5$. $W$ and $E$ stand for warping
 and extrapolation, respectively.}
	\label{fig:flow}

\end{figure}

Given two rendered frames $F_t$ and $F_{t+1}$, our goal is to insert $n$ new frames between these two frames without
using $F_{t+1}$, where $n$ $\in$ $[1,3]$. We represent these frames as $P_i$, where $i$ falls within the interval
$[1,3]$ and the frame $P_i$ is displayed on the screen at time $t+(i/4)$.  We have two options: warping and
extrapolation.  For the ensuing discussion, please refer to
Figure~\ref{fig:flow}. We need to make a decision about which frame to display -- warped or extrapolated at three time
instants: $t+0.25$, $t+0.5$, $t+0.75$. We refer to these frames as $P_1$, $P_2$, and $P_3$,
respectively.\\

\textbf{Decision at $\mathbf{t}$}: The decision ($d_1$) at this time decides which frame to display at $t+0.25$: the
warped version of $F_t$ ($=W(F_t)$) or the last displayed frame $F_t$. The second choice would also result in the
extrapolated $F_t$ ($=E(F_t)$) being displayed at $t+0.5$. This is because the extrapolated frame would only be
available at $t+0.5$.

\textbf{Decision at $\mathbf{t+0.25}$}:  The decision at this time instant decides which frame to display at $t+0.5$. If the
outcome of $d_1$ was to extrapolate, we do not need to take any decision at this point since we display $E(F_t)$ at
$t+0.5$ as explained above. If $d_1$ chose warping, then at this point we make decision $d_2$, which chooses to display
either the warped of the last displayed frame, i.e., $W(P_1)$ or the extrapolated version of $F_t$ ($E(F_t)$).  

\textbf{Decision at $\mathbf{t+0.5}$}: We decide which frame to display at $t+0.75$. Depending
on the outcomes of the previous two decisions, there are three possible cases. First, we consider the case where $d_1$ chose
extrapolation. In this case, the decision $d_3$ decides whether to display the warped version of $P_2$ ($W(P_2)$) or just
display $P_2$ since extrapolation will not be able to generate the frame by $t+0.75$.  The other two cases are relevant
if $d_1$ chose warping. For both of these cases, decisions $d_4$ and $d_5$ choose between the warped versions
of $P_2$ ($W(P_2)$) and
extrapolated $P_1$ ($E(P_1)$). However, the frames $P_1$ and $P_2$ themselves would depend on the
decision taken at $d_2$.
Based on these decisions, six scenarios or {\em decision paths} are possible (shown in Table~\ref{tab:pos_sce}).

\begin{table}[!htb]
\footnotesize
\begin{center}
\resizebox{0.99\columnwidth}{!}{
\begin{tabular}{| l |l | l| l| l|} 

\hline
\textbf{Scenario} &
$\mathbf{P_1}$ & $\mathbf{P_2}$ & $\mathbf{P_3}$ & \textbf{Frame rate}\\
&  &  & & \textbf{upsampling} \\ 
\hline\hline
$\mathbf{S_1}$ & $F_1$  & Extrapolated $F_1$  & Extrapolated $F_1$ & 2$\times$\\
(ExtraNet) &  (No new frame)  &  &  (No new frame) &   \\ \hline
$\mathbf{S_2}$ & $F_1$   & Extrapolated $F_1$  & Warped $P_2$ & 3$\times$ \\
 & (No new frame) &  &  &   \\ \hline
$\mathbf{S_3}$ & Warped $F_1$ & Extrapolated $F_1$ & Extrapolated $P_1$ & 4$\times$\\  \hline
$\mathbf{S_4}$ & Warped $F_1$ & Extrapolated $F_1$ & Warped $P_2$ & 4$\times$\\  \hline
$\mathbf{S_5}$ & Warped $F_1$ & Warped $P_1$ & Extrapolated $P_1$ & 4$\times$\\  \hline
$\mathbf{S_6}$ & Warped $F_1$ & Warped $P_1$ & Warped $P_2$ & 4$\times$\\
\hline
\end{tabular}
}
\end{center}
\caption{Possible scenarios (decision paths) based on the type of synthesized frames- $P_1$, $P_2$, and $P_3$}
\label{tab:pos_sce}
\vspace{-6mm}
\end{table}

\subsection{RL-based Decision Predictor}
\label{rlmodel}

As mentioned in Section~\ref{sec:Introduction}, we propose an RL-based model~\ref{RL} that intelligently takes the
decisions shown in Figure~\ref{fig:flow} to provide the best overall performance both in terms of quality as well
as frame rate. RL-based approaches are germane to this scenario because we take decisions in a potentially uncertain
environment, and that too with partial information.  

In this section, we define the structure, features, and parameters used in our network. As discussed in
Section~\ref{staterep}, we feed the state information of the scenes to the model. The motivation behind the choice of
values taken to represent the state has been summarized in Section ~\ref{feature}. To better understand the complexity
of the current scene, we use the past three states with the current state as the input to make a prediction. Each state
is defined by a tuple of vectors: the environment vector $E_t$ (Eqn. \ref{eqn:et}) and the temporal vector $T_t$.
The environment vector
$E_t$ for any frame $F_t$ contains the values of the features mentioned in Table~\ref{features}. We also consider
the variance in the horizontal and vertical direction of the motion vector separately. We also use the rendering
resolution ($R$) of $F_t$ as a single integer, $R = R_h \times R_w$ (horiz$\times$ vert).

\begin{equation}
\label{eqn:et}
	E_t=[ N_d, EMD_{W_n}, EMD_{W_p}, Var_x(F_{mv}), Var_y(F_{mv}), R ]
\end{equation}

To encode information about the current decision, we use another vector $T_t$ of length five, which represents the five
states in Figure~\ref{fig:flow} as a one-hot encoded vector. The state is represented as $S_t = (E_t,T_t)$. This tuple
is flattened into a single vector and the concatenation of all the 4 state vectors (current and last three frames)
are fed into the RL network. The length
of the vector is 44 (11$\times$4) 4-byte floats (represented in fixed point). The model gives an output vector of length two, which corresponds to the rewards
associated with the two choices: warping and extrapolation. The choice that gives the maximum reward is chosen
finally. Hence, the network is a mapping defined by $EW_{net}: \mathbb{R}^{44} \longrightarrow \mathbb{R}^2 $. The
predicted choice by the model at time $t$ ($A_t$) is calculated as:

\begin{equation}
	A_t=\arg_{i} max(EW_{net}([S_t,S_{t-1},S_{t-2},S{t-3}]))
\end{equation}

To train our network to potentially foresee the future frames with high variation in features, we
minimize the error between the predicted reward and the weighted sum of the current reward and the reward associated
with the next best action. This ensures that the model is able to predict future high-variance sequences and take
appropriate decisions. When the scene has less variation the model, we should prefer warping whereas when the scene
is highly dynamic, the model needs to predict that and migrate towards extrapolation. It is important for our model to anticipate the
future as the immediate best action might not be the best action (the local optima may not be global).

\subsubsection{Reward Function}

The reward function is the sum of the below at each decision point. Here, $MSE$ refers to the mean square error.
\begin{equation}
\begin{split}
& R = \Delta PSNR + \Delta SSIM + \alpha \\
& PSNR = 10 \times log_{10} \left ( 255^2/MSE \right ) \\
& SSIM = \text{Structural similarity between two images}
\end{split}
\end{equation}

\begin{itemize}
    \item The gain/loss in PSNR~\cite{psnr} and SSIM~\cite{psnr} due to an {\em action}, which is the difference between the PSNR/SSIM of the frame
generated due to the chosen action and the PSNR/SSIM of the frame that would have been generated by the other action. 
These metrics use the ground truth as the baseline.
	\item $\alpha$ = -0.1, loss associated with dropping frames 
\end{itemize}

The network estimates the reward function $R(S_t,A_t;\theta_i)$ at state $S_t$
for an action $A_t$ with the network parameters $\theta_i$ at the $i^{th}$ training step. The function
$R(S_t,A_t;\theta_i)$ is the maximum reward in $EW_{net}([S_t,S_{t-1},S_{t-2},S{t-3}])$.  Our loss can be defined as:

\begin{equation}
	L_i = \mathbb{E}\left[ \left( r+\gamma \max_{A_{t+1}}R(S_{t+1},A_{t+1};\theta_{i-1})-R(S_t,A_t;\theta_{i}) \right) ^2 \right]
\end{equation}

where $\gamma$ (=0.95) is the discount factor, which is used to tune the importance the model gives to future moves and $r$
is the ground truth reward at that point. 

Given that our input size is small,
we can afford a network with three fully-connected+ReLU layers ($44\times 128$, $128\times 256$, $256\times 128$,
$128\times 2$) 
followed by one output layer ($2\times1$). 
Our network is trained with 3000 data points and tested with 1000 data points per benchmark. We
generate the frames at 30 fps. We employ LOOCV cross-validation to prevent overfitting: test data points corresponding to one
benchmark and use the remaining data points for training. We repeat this procedure for each benchmark (essentially, we
rotate the train-test set) and report the mean.  

\subsubsection{Hardware Implementation of the Predictor}
\begin{figure}[!htbp]
	\centering
	\includegraphics[width=0.99\columnwidth]{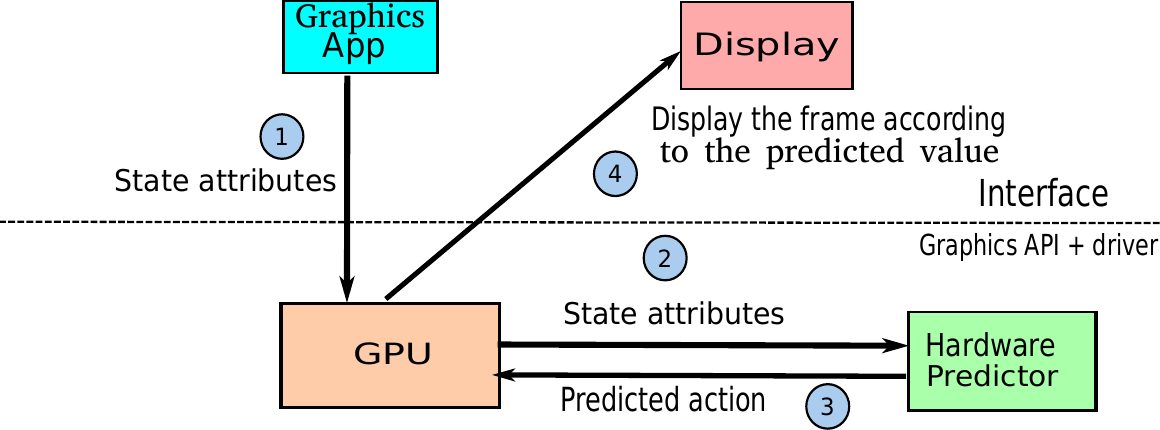}
	\caption{System architecture}
	\label{arch}
\vspace{-6mm}
\end{figure}

We train the
predictor offline and use it for online inferencing. 
We implemented the predictor in Verilog and synthesized it. We used 4 simple cores in our design
given its simplicity: each core has a four-stage pipelined
architecture. Since, most of the operations in neural network
inferencing are matrix multiplication and addition, we designed ten 8-bit multiply-accumulate (MAC) units along with one
adder and one multiplier unit for each core. Each core has a private cache of 16 KB. The system architecture of the
proposed system is shown in figure~\ref{arch}.
  
\section{Results and Analysis}
\label{sec:Evaluation}

\subsection{Performance Analysis}

\subsubsection{Comparison of Various Possible Decision Paths}
\label{perf_with_scen}

We showed the possible decision paths in Table~\ref{tab:pos_sce}. In this section,
we compare the quality of the generated scenes.
The results are shown in
Table~\ref{perf_scenario}. We make the following observations from the results:\\ 

\circled{1} For most of the
benchmarks, \exwarp performs the best. Even if it is not the best, the values are comparable with the rest of the
methods. \\ 
\circled{2} There is an 18.02\% and 6.58\% increase in PSNR and SSIM in \exwarp, respectively, as compared to  $S_1$ --
pure extrapolation-based method. \\ 
\circled{3} When compared to $S_6$, a pure warping-based method, there is a 9.24\%
and 0.04\% increase in PSNR and SSIM, respectively.

\begin{table}[!htb]
\footnotesize
\begin{center}
	
\resizebox{0.99\columnwidth}{!}{
  \begin{tabular}{|l|l|l|l|l|l|l|l|l|}
    \hline
 \multicolumn{2}{|c|}{ \multirow{2}{*}{\textbf{Scenes}}}  &\multicolumn{6}{c|}{\textbf{Scenarios}} &  \multirow{2}{*}{\textbf{\exwarp}}\\
   
 \multicolumn{2}{|c|}{}   &   $\mathbf{S_1}$ & $\mathbf{S_2}$ & $\mathbf{S_3}$ & $\mathbf{S_4}$ & $\mathbf{S_5}$ & $\mathbf{S_6}$ & \\
    \hline
 \multirow{7}{*}{\rotatebox[origin=c]{90}{\textbf{PSNR(dB)}}}  
   &  LB &  20.63  & 20.58 & 20.05  & 23.91 & 21.43 & 25.01 & \textbf{28.65}\\
    \cline{2-9}
   & TR &19.82 & 19.18 & 19.11  & 22.60 & 20.19 & \textbf{23.68} & 17.30\\
   \cline{2-9}
    & VL & 20.33 & 19.92 &  24.90  & 36.36 & 32.68 & 43.70 & \textbf{44.66}\\
    \cline{2-9}
  &  TN &  18.29 & 17.56 &  16.94 & 19.71 & 19.97 & 17.39 & \textbf{24.10}\\
   
  & TN2 &   \textbf{18.70} & 16.85 & 15.60   & 17.98 & 17.35 & 15.06 & 17.12\\
  
  & TN3 &  \textbf{19.26}  & 17.14 &  15.64  & 17.91 & 16.98 & 14.79 & 17.01\\
  \cline{2-9}
  &  SL &  29.06 & 28.86 & 30.11  & 31.38 & 32.10 & 30.93 & \textbf{32.62} \\
   
  & SL2 &  31.17 & 29.20 &  28.20  & 29.97 & 26.83 & 27.59 & \textbf{32.33}\\
  
  & SL3 &  30.99 & 29.40 & 27.83 & 28.52 & 26.43 & 26.90 & \textbf{32.03}\\
  \hline
  \hline
    \multirow{7}{*}{\rotatebox[origin=c]{90}{\textbf{SSIM}}} 
    &  LB & 0.64 & 0.63 & 0.55 & 0.76 & 0.60 & 0.81 & \textbf{0.87}\\
    \cline{2-9}
   & TR &  0.63 & 0.59 & 0.51 & \textbf{0.69} & 0.54 & 0.71 & 0.55\\
   \cline{2-9}
    & VL &  0.80 & 0.77 & 0.69  & 0.92 & 0.75 & 0.99 & \textbf{0.99} \\
    \cline{2-9}
  &  TN &  0.75 & 0.73 & 0.64  & 0.78 & 0.79 & 0.65 & \textbf{0.81}\\
   
  & TN2 &  \textbf{0.78} & 0.71  & 0.61  & 0.70 & 0.66 & 0.56 & 0.69\\
  
  & TN3 &  \textbf{0.82} & 0.74 &  0.62  & 0.72 & 0.66 & 0.57 & 0.71\\
  \cline{2-9}
  &  SL  & 0.88 & 0.87 & 0.77 & 0.92 & \textbf{0.94} & 0.78 & 0.93\\
   
  & SL2  & 0.88 & 0.85 &  0.72  & \textbf{0.88} & 0.71 & 0.87 & 0.87 \\
  
  & SL3  & \textbf{0.88}  & 0.86 &   0.72 & 0.87 & 0.72 & 0.87 & 0.87 \\
  \hline
  \end{tabular}
  }
 \end{center}
 \caption{Performance comparison across all scenarios (decision paths)}
\label{perf_scenario}
\vspace{-6mm}
\end{table}

\subsubsection{Comparison with the State-of-the-Art}
\label{perf_inter}
In this section, we compare the performance of our proposed model with two interpolation-based methods and 
ExtraNet. The interpolation-based methods are Softmax Splatting~\cite{softmax} and EMA-VFI~\cite{EMA}. 
All three methods are DNN-based techniques. Softmax splatting uses forward warping; it uses forward and backward motion
flow (reprojection). However, in this approach, multiple pixels may map to the same
target location in frame $F_t$. Softmax splatting uses a modified softmax layer, which takes the frame's depth data to
handle this ambiguity. EMA-VFI uses a transformer network to perform frame interpolation. We show the performance of
these methods in Table~\ref{perf_inter_extra}. For PSNR, \exwarp is the best for 4/9 benchmarks and the
second best for two. There is a large difference only in the case of LB and TR. For the SSIM metric, both ExtraNet and
\exwarp do well.

\begin{table}[!h]
\footnotesize
\begin{center}
	
\resizebox{0.99\columnwidth}{!}{
  \begin{tabular}{|l|l|l|l|l|l|l|}
    \hline
 \multicolumn{2}{|c|}{ \multirow{2}{*}{\textbf{Scenes}}}  &\multicolumn{2}{c|}{\textbf{Interpolation}} & \multicolumn{2}{c|}{\textbf{Extrapolation}} \\
   
 \multicolumn{2}{|c|}{}    &\textbf{EMA-VFI} &   \textbf{Softmax Splatting} & \textbf{ExtraNet} & \textbf{\textit{\exwarp}}  \\
    \hline
 \multirow{7}{*}{\rotatebox[origin=c]{90}{\textbf{PSNR(dB)}}}  
   &  LB  & \textbf{49.52}  & 48.74 & 20.63  &  28.65\\
    \cline{2-6}
   & TR &  \textbf{24.60} & 23.42 & 19.82 &  17.30 \\
   \cline{2-6}
    & VL &20.86 & 20.54  &  20.33  & \textbf{44.66} \\
    \cline{2-6}
  &  TN &  14.40 & 13.84 & 18.29  & \textbf{24.11} \\
   
  & TN2 &  13.42 & 13.47 &   \textbf{18.70} &  17.12 \\
  
  & TN3 &  14.15 & 14.53 &  \textbf{19.27}  & 17.01 \\
  \cline{2-6}
  &  SL &  28.57 & 24.07 & 29.06  &  \textbf{32.62}\\
   
  & SL2 &  24.58 & 22.74 &  31.16  & \textbf{32.33} \\
  
  & SL3 &  32.53 & \textbf{34.95} &  30.99  & 32.03 \\
  \hline
  \hline
    \multirow{7}{*}{\rotatebox[origin=c]{90}{\textbf{SSIM}}} 
    &  LB & \textbf{0.99}  & \textbf{0.99}  & 0.64 & 0.87 \\
    \cline{2-6}
   & TR &  \textbf{0.97} & 0.95 & 0.63 & 0.55  \\
   \cline{2-6}
    & VL &  0.94 & 0.93 & 0.80   & \textbf{0.99} \\
    \cline{2-6}
  &  TN & \textbf{0.82} & 0.77 & 0.75  & 0.81 \\
   
  & TN2 &  0.71 & 0.59 &   \textbf{0.78} & 0.69 \\
  
  & TN3 &  0.75 & 0.64  &  \textbf{0.82}  & 0.71 \\
  \cline{2-6}
  &  SL &  0.61 & 0.26 &  0.88 & \textbf{0.93} \\
   
  & SL2 &  0.40  & 0.27 & \textbf{0.88}   & 0.87 \\
  
  & SL3 &  0.75 & 0.83 &  \textbf{0.88}  & 0.87 \\
  \hline
  \end{tabular}
  }
 \end{center}
 \caption{Performance comparison with the state-of-the-art}
\label{perf_inter_extra}
\vspace{-6mm}
\end{table}

\subsection{Frame rate (FPS)}
\begin{figure}[!htb]
	\centering
	\includegraphics[width=0.99\columnwidth]{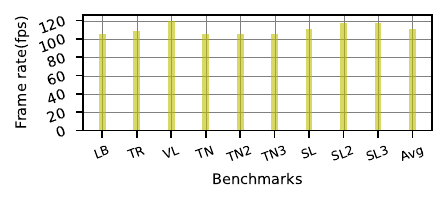}
	\caption{FPS}
	\label{fps}
\vspace{-6mm}
\end{figure}

In this section, we plot the final frame rate achieved using \exwarp for each benchmark. As mentioned in
Section~\ref{rlmodel}, the original frame rate was 30 fps. The results are shown in Figure~\ref{fps}. The insights from
the results are as follows:\\ 

\circled{1} The effective upsampled frame rate for all the benchmarks is more than 100 fps. We compute this based on the
number of new frames that we actually manage to insert. The more we extrapolate, lower is this figure. \\
\circled{2} The average frame rate across benchmarks is almost 110 fps, hence the supersampling factor is nearly 4 for
our proposed method. Note that this is more than all state-of-the-art work.

\subsection{Warping vs Extrapolation}
\begin{figure}[!htb]
	\centering
	\includegraphics[width=0.99\columnwidth]{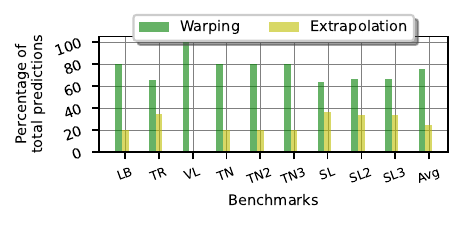}
	\caption{Breakup of the predictions made by the predictor}
	\label{actions}
\vspace{-6mm}
\end{figure}

Our proposed model, \exwarp, predicts the best method between warping and extrapolation. In this section, we plot the
prediction pattern of our predictor. The results are shown in Figure~\ref{actions}. The observations from the results
are as follows:\\ \circled{1} For most of the benchmarks, the ratio between warping and extrapolation is 80:20 except
\textit{VIL}.\\ \circled{2} The average percentage for warping across benchmarks is 75.86\%.

\subsection{Synthesis Results}
We used the popular tool NNGen~\cite{nngen} to generate a baseline Verilog code for our neural network.
We then made modifications to it and manually tuned it.  We used
the Cadence Genus Tool (TSMC 28 nm technology) to synthesize the design and obtain the power, area and timing numbers.
Table~\ref{pred_lat} shows the area and power overheads of the hardware predictor. The total area is 0.12 $mm^2$, which
is negligible. Also, the latency, 6.2 ns, is insignificant.

\begin{table}[!h]
\vspace{-1mm}
\footnotesize
\begin{center}
	
\resizebox{0.99\columnwidth}{!}{	
\begin{tabular}{ |l|p{45mm}|l|} 

\hline
\rowcolor{gray}
\textbf{Parameter} &
\textbf{Value} \\ 
\hline\hline
Tool & Cadence Genus, 28 $nm$\\
\hline
Area & 0.12 $mm^2$\\
\hline
Power & 9.12 $mW$  \\
\hline
Latency & 6.2 $ns$ \\
\hline

\end{tabular}
}
\end{center}
\caption{Overheads of the hardware predictor}
\label{pred_lat}
\vspace{-4mm}
\end{table}

\begin{table*}[!htb]
	\footnotesize
	\begin{center}
\resizebox{0.99\textwidth}{!}{
		\begin{tabular}{ |l|l|l|l|l|l|l|} 
			
			\hline
			\rowcolor{gray}
			\textbf{Year} &
			\textbf{Work} &
			\textbf{Coherence Exploited} &
			\textbf{Method Used} &
			\textbf{ML-based} &
			\textbf{Upsampling}\\ [0.5ex] 
			\hline\hline
			2007 & Nehab et al.~\cite{nehab2007accelerating} & Spatial and temporal  & Interpolation & \textcolor{red}{$\times$} & \\
			\hline
			2010 & Andreev et al.~\cite{andreev2010real} & Temporal  & Interpolation & \textcolor{red}{$\times$} & x to 60 fps \\
			\hline
			2010 & Didyk et al.~\cite{didyk2010perceptually} & Temporal  & Interpolation & \textcolor{red}{$\times$} & 40 fps to 120 fps \\
			\hline
			2010 & Herzog et al.~\cite{herzog2010spatio} & Spatial and temporal  & Interpolation & \textcolor{red}{$\times$} & \\
			\hline
			2011 & Yang et al.~\cite{yang2011image} & Temporal & Interpolation & \textcolor{red}{$\times$}  & \\
			\hline
			2012 & Bowles et al.~\cite{bowles2012iterative} & Temporal  & Interpolation & \textcolor{red}{$\times$} & \\
			\hline
			2018 & SAS~\cite{mueller2018shading} & Temporal  & Interpolation & \textcolor{red}{$\times$} & from (7.5,15,30, 60) fps to 120 fps \\
			\hline
			2021 & ExtraNet~\cite{extranet} & Temporal  & Extrapolation & \textcolor{green}{\checkmark} & upto 2$\times$ (30 fps to 60 fps) \\
			\hline
			2022 &  DLSS 3~\cite{DLSS3} & Spatial and temporal   & Interpolation & \textcolor{green}{\checkmark}  & upto 4$\times$\\
			\hline
			\textbf{2023} & \textbf{Our work} & \textbf{Temporal}  & \textbf{Extrapolation} & \textbf{\textcolor{green}{\checkmark}} & \textbf{upto 3$\times$} \\
			\hline

		\end{tabular}
}
	\end{center}
	\caption{A comparison of related work}
	\label{table:relwork}
\vskip -6mm
\end{table*}

\section{Related Work}
\label{sec:RelatedWork}

Over the past few years, a variety of solutions have been developed that exploit the spatial and temporal coherence
present in graphics applications to increase the frame rate of graphics applications and synchronize the GPU
refresh rate with the display refresh rate for a seamless user experience. Recent works primarily focus on \circled{1}
predicting new frames using interpolation~\cite{yang2011image,nehab2007accelerating,herzog2010spatio, andreev2010real,
DLSS2} and \circled{2} generating new frames using extrapolation~\cite{extranet} to increase the frame rate. We present
a brief comparison of related work in Table \ref{table:relwork}. The high processing cost per frame is the primary cause
of the low frame rate~\cite{nehab2007accelerating}, and previous works aim to decrease this cost. According to Herzog et
al.~\cite{herzog2010spatio}, the visual appearance, illumination parameters, etc., are nearly identical between any two
consecutive frames and sometimes within a single frame -- they are
 known as temporal and spatial coherence, respectively. These effects can be
exploited to reduce the overall processing cost per frame. 
They mention that although approaches based on reducing the resolution of a frame, predicting the next few frames, and then
performing
spatial supersampling are very efficient, such approaches
can also undersample or blur sharp image features such as edges quite frequently. On the
other hand, pure temporal supersampling is markedly better. This work is based on exclusive temporal
supersampling. 

\subsection{Interpolation}
Previous approaches \cite{nehab2007accelerating} that use the temporal supersampling to fill in a frame between a pair
of rendered frames use the interpolation process that is guided by the scene flow: the 3D velocities of visible surface
points between two frames. For each pixel in an intermediate frame, the motion vector indicates where to pull pixel
information from the original frames. Early approaches had a fundamental drawback, which was that
whenever the scene
contained regions that were visible in the current frame but were not in the previous one, the results
were sub-par. Although Bowles et
al. \cite{bowles2012iterative} proposed an efficient way to fix this using an iterative method called fixed point
iteration (FPI), this  did not provide satisfactory results. To handle this case, various works
~\cite{yang2011image,didyk2010perceptually, DLSS3, mueller2018shading} propose  a bidirectional reprojection method that
temporally upsamples rendered content by reusing data from both the backward and forward temporal directions. Didyk et
al.~\cite{didyk2010perceptually} use motion flow to warp the previously shaded result into an in-between frame that is
then locally blurred to hide artifacts caused by morphing failures. Finally, they compensate for the lost
high-frequencies due to this blur by adding additional high frequencies wherever necessary. They perform an upsampling
from 40 Hz to 120 Hz. Similarly, NVIDIA's DLSS 3~\cite{DLSS3} use the optical flow computed from both the backward and
temporal directions to interpolate the frame. DLSS3 has two major components: an optical-flow generator and a frame
generator apart from the supersampling network. They use the in-built accelerator in their latest GPU architecture Ada
for the optical flow generation. The frame generator uses an AI-accelerated network that takes the computed optical flow
to generate an entirely new frame. As shown in Figure~\ref{inter_vs_extra}, this approach increases the frame rate but
also leads to an increased input latency that can easily be perceived by users. Since our approach is not based on 
optical flow fields, it does not require future frames to predict a new frame. 

Nehab et al.~\cite{nehab2007accelerating} use a reverse reprojection-based caching technique to store the information
that can be reused in the next frame, thereby avoiding the recomputation of the entire frame.  
Andreev et al.~\cite{andreev2010real} propose an approach to maintain a
consistent rate of 60 fps by dividing a frame into two parts: slow-moving and
fast-moving, and rendering each one at a different rate (slower parts at a lower rate and faster parts at a higher
rate). This approach works because some tests have shown that the temporal coherence of slowly moving parts is greater
than that of other parts.  Such approaches increase
the time required for an application to construct a frame while maintaining a
constant frame rate. 

\subsection{Extrapolation}

This is a very sparse area of research. The only prominent work that we are aware of is ExtraNet~\cite{extranet}.
This was discussed in detail in Section~\ref{sec:extra_lat}.

\section{Conclusion}
\label{sec:Conclusion}
With high-frequency displays becoming increasingly popular, it is now necessary to generate
frames for real-time applications at higher rates. Since applications are very demanding in terms of processing power,
it is not possible for even the most capable GPUs to constantly provide a high frame rate at an HD/4k resolution.
It has become evident that new frames need to be generated without having to go through the entire graphics pipeline.
This work illustrates one such method, \exwarp, of supersampling in the temporal domain, while maintaining frame quality.  The
existing methods use extrapolation to increase the frame rate but we observed that it is not always necessary to use an
expensive method like extrapolation and that the decision to extrapolate or use a faster method such as warping can be
intelligently made. We designed such a predictor to take this decision.
We were able to achieve nearly four times the frame rate ($\approx 120$ Hz) with a reasonably small reduction in the
quality.


\bibliographystyle{IEEEtranS}
\bibliography{references}

\end{document}